\documentclass[a4paper,twoside]{article}

\newif\ifanonymous
\anonymousfalse

\usepackage[T1]{fontenc}
\usepackage[utf8]{inputenc}
\usepackage{epsfig}
\usepackage{graphicx}
\usepackage{subcaption}
\usepackage{calc}
\usepackage{amssymb}
\usepackage{amstext}
\usepackage{amsmath}
\usepackage{amsthm}
\usepackage{mathtools}
\usepackage{multicol}
\usepackage{pslatex}
\usepackage{apalike}
\usepackage{algorithm2e}
\usepackage[bottom]{footmisc}
\usepackage{booktabs}
\usepackage{xcolor}
\usepackage{comment}
\usepackage{tabularx}
\usepackage{url}
\usepackage{enumitem}
\usepackage{SCITEPRESS} 

\begin{document}

\title{Strategic and Robust Task Allocation for Combat Objectives in Adversarial Multi-Agent Environments}

\ifanonymous
\author{}
\else
\author{
\authorname{Jonas Pech\'e\sup{1}\orcidAuthor{0009-0001-8250-9008}, Dmitry Matveev\sup{2}, Vladimir Ciric\sup{3}, Lizaveta Sakalova\sup{4} and G\"unter Wallner\sup{1}}
\affiliation{\sup{1}Johannes Kepler University Linz, Computer Graphics, Linz, Austria}
\affiliation{\sup{2}Wargaming, Nicosia, Cyprus}
\affiliation{\sup{3}Wargaming, Belgrade, Serbia}
\affiliation{\sup{4}Wargaming, Warsaw, Poland}
\email{j\_peche@wargaming.net, d\_matveev@wargaming.net, v\_ciric@wargaming.net, l\_sakalova@wargaming.net, guenter.wallner@jku.at}
}
\fi


\abstract{Task allocation in contested-objective team games requires coordinated decisions under uncertainty, spatial constraints, and changing tactical conditions. In such settings, multiple agents must anticipate both teammate behavior and opponent actions. We study a hybrid predict-then-optimize approach for assigning controllable agents to objectives in real time. A masked multi-head model predicts tactical signals for each agent, including survival chance, local superiority, combat interaction, travel time, and objective contribution. These signals are combined with rule-based terms in a score matrix, and a minimum-cost flow solver computes coordinated global assignments constrained by capacity, occupancy, stability, and defensive strategies. Learning is limited to tactical scoring, while explicit rules and solver structure preserve controllability and interpretability. We evaluate the system using predictive metrics, offline assignment diagnostics, online A/B evaluation, and latency measurements. The system is an R\&D research prototype built on a commercial game and is not deployed in the shipped game. The results indicate that learned tactical estimates can support stable, feasible, and inspectable objective allocation under strict real-time constraints.}

\onecolumn \maketitle \normalsize \setcounter{footnote}{0} \vfill

\section{\uppercase{Introduction}}

In many team-based adversarial games, players compete over spatial objectives that must be captured, defended, or contested. Teams win not only by eliminating opponents, but also by controlling objectives that generate score or other strategic advantages. A central decision problem in such games is therefore to determine which agents should move to which objectives and how limited team resources should be distributed across the map. However, task allocation in multi-agent environments is difficult because agents must be assigned to tasks under utility, resource, and coordination constraints~\cite{korsah2013taxonomy,gerkey2004formal}. The problem becomes especially challenging when the value of objectives is dynamic, contested, and strongly influenced by other agents. In objective-control games, where teams compete over shared objectives, assigning one agent to a capture objective not only changes its own impact, but also the value of other agents through force concentration, exposure, and shifts in opportunities. Similar inter-agent dependencies are also central in multi-robot task allocation and market-based coordination, where local utility estimates alone are often insufficient for globally coherent team behavior~\cite{dias2006market,choi2009cbba}.

Artificial intelligence (AI) for commercial games often uses hand-authored heuristics for capture and defense decisions. While these systems are fast, they are difficult to maintain across map changes, rule updates, and evolving player behavior. Classical assignment methods provide efficient solutions when utilities are fixed and tasks are independent~\cite{kuhn1955hungarian}, but they do not encode many-to-one allocation, objective capacities, human occupancy, or game-specific stability constraints out of the box. Network-flow 
formulations extend this view to capacities and richer feasibility constraints~\cite{bertsekas1998network,goldberg1990mincost}. However, local game-AI heuristics can fail in contested settings because they favor short-term value over global coverage, force balance, and simultaneous objective pressure. This motivates methods that maintain the speed and controllability of heuristics while also adding global coordination over capacity, force balance, and simultaneous capture objectives.

In this paper we study the application of a learning-assisted constrained optimization pipeline to strategic task allocation in team-based contested-objective battle. This work and described system is an internal research and development (R\&D) project and is not part of the released game. As a use case we focus on the game \emph{World of Tanks: HEAT}~\cite{wargaming_wot_heat_2026}, a team-based vehicle shooter in which two teams compete for certain objectives. In particular, we study whether this pipeline can produce coordinated, stable, and real-time feasible recommendations. The primary strategic outcome is match win rate, and a recommendation is considered strategic if it improves long-term win probability rather than producing a locally attractive move only. The central comparison is against the game's built-in closest-base AI baseline, using offline evaluation on logged battles and battle-level outcome metrics where available.

Thus, the specific research challenge addressed in this paper is to assign every currently controllable bot to an actionable contested objective every few seconds, despite controlling only part of the allied team, while respecting objective capacities, current human occupancy, tactical uncertainty, and the need for stable team-level coordination. Each decision must be produced from the current battle snapshot, remain inspectable and designer-controllable, and fit comfortably within the game's real-time latency budget. By \emph{robust} we mean stable, feasible assignments under rapidly changing battle states, not robustness to input noise. In summary, the main contributions of this paper are:
\begin{enumerate}
    \item We formulate contested objective allocation in adversarial multiplayer battles as a real-time team-coordinated assignment problem under partial controllability.
    \item We implement a hybrid framework combining multi-head tactical outcome prediction with constrained minimum-cost flow optimization.
    \item We use robustness-oriented constraints and penalties to improve assignment stability and feasibility under rapidly changing battle conditions.
    \item We evaluate the resulting recommender using predictive metrics, offline assignment diagnostics, online evaluation, and latency measurements.
\end{enumerate}

\section{\uppercase{Related Work}}

Our problem relates to classical assignment and network optimization, multi-robot task allocation, industrial dispatch, learning-enhanced combinatorial optimization, as well as game AI.

\subsection{Assignment, Flow, and Task Allocation}

For standard one-to-one assignment problems, bipartite matching under linear costs is commonly solved using the Hungarian method~\cite{kuhn1955hungarian}. For problems closer to min-cost flow and minimum-cost circulation, general network optimization treatments provide the theoretical basis~\cite{bertsekas1998network}, while cost-scaling and successive-approximation approaches provide efficient practical algorithms~\cite{goldberg1990mincost}. These results motivate using min-cost flow as the final decision layer in settings where global feasibility and capacity constraints must be enforced.

Early multi-robot task allocation work showed that many coordination problems can be viewed as assigning tasks to agents under resource and utility constraints, while emphasizing that different task structures require different algorithmic regimes~\cite{gerkey2004formal,korsah2013taxonomy}. This perspective is directly relevant to tactical assignment in games, where units such as vehicles differ by role, health, mobility, and location, and where tasks such as capturing, defending, escorting, or recovering are connected. A survey on market-based coordination~\cite{dias2006market} showed that action- and utility-based mechanisms can produce scalable allocation behavior in multi-agent systems. Decentralized auction methods such as the \emph{Consensus-Based Bundle Algorithm} provide robust task allocation under local communication and partial decentralization~\cite{choi2009cbba}. In contrast, our setting permits centralized coordination, as such we use min-cost flow to prioritize globally feasible and controllable assignments.

\subsection{Industrial Dispatch and Fleet Coordination}

Ride-hailing work treats online driver-order assignment as a sequential decision problem where local matching can be suboptimal because it ignores future supply-demand dynamics. For example, predictive modeling has been combined with coordinated assignment~\cite{xu2018ridesharing}; deep value estimates have been used in production dispatch~\cite{tang2019deepvalue}; and ridesharing dispatch has been modeled with mean-field multi-agent reinforcement learning~\cite{li2019meanfield}. These works show that allocation quality depends on the collective state induced by all assignments, not only on isolated assignment scores.

Work on warehouse coordination~\cite{wurman2008kiva}, on the other hand, describes a system for coordinating hundreds of autonomous mobile robots in warehouse environments by jointly considering task allocation, path planning, congestion management, and timing. Even though game maps are less structured than warehouse floors, the same observations apply: the quality of an assignment depends on travel time, route interference, and whether multiple agents are sent into the same limited region without sufficient marginal gain.

Industrial delivery work~\cite{zhu2020ofct} further highlights the importance of time modeling and shows that end-to-end timing is central in dispatch pipelines such as for order fulfillment cycle time estimation in food delivery. In tactical games, ignoring travel and completion time misprices delayed movement between objectives, late returns to defend objectives, and opportunities that disappear as the match state changes (see Section~\ref{sec:environment}). These systems motivate tactical value estimates that include delayed effects, timing uncertainty, and opportunity cost.

\subsection{Learning for Downstream Optimization}

Another important direction is learning for downstream optimization rather than for standalone prediction accuracy. So-called predict-then-optimize pipelines predict optimization parameters, scores, or probabilities from observed features, and then solve the optimization problem~\cite{elmachtoub2022spo,elbalghiti2019generalization}. However, prior work~\cite{elmachtoub2022spo} points out that this separation can be suboptimal because small prediction errors may induce large decision errors, while some apparently large prediction errors may have no effect on the final decision. It thus proposes the \emph{Smart Predict-then-Optimize} framework, which introduces loss functions aligned with downstream decision quality. For tactical assignment in games, this means that per-agent predictions should be judged not only by calibration or accuracy, but also by the assignments induced by the solver.

Neural heuristics for combinatorial optimization problems provide a related direction. Attention-based routing work~\cite{kool2019attention} shows that attention models can learn competitive policies for routing problems. This suggests that parts of the assignment pipeline could be learned from training data, but also that specialized solvers remain strong when the optimization structure is known. We hence evaluate predictions not only by accuracy but also by the assignments induced downstream.

\subsection{Game AI and Tactical Allocation}

Real-time strategy (RTS) games have frequently been used as benchmarks for game AI because they combine adversarial decision making, partial observability, spatial reasoning, real-time control, and tactical--strategic abstraction~\cite{ontanon2013survey,robertson2014review}. Existing work has studied tactical coordination through methods such as influence maps, potential fields, search, evolutionary optimization, and machine-learning and reinforcement-learning methods for unit control and combat micromanagement (e.g.,~\cite{ontanon2013survey,robertson2014review}). For example, co-evolved influence maps have been used for spatial team tactics in an RTS setting~\cite{avery2010coevolving}, while task allocation in an RTS game has been studied using a swarm-intelligence approach whose parameters are tuned by a genetic algorithm~\cite{tavares2014swarmrts}.

However, applying these methods in our setting requires handling small-team adversarial objectives, mixed human--bot teams with partial controllability, strict real-time replanning, and learned tactical scoring coupled to a constrained solver. We therefore study tactical allocation in a competitive multiplayer mode with strict latency and deployment requirements for stability, interpretability, and controllability.

\section{\uppercase{Methodology}}

Our proposed task allocation framework consists of a sequence of subsystems that map a given game situation into suitable task allocations. In the following, we first describe the environment (Section~\ref{sec:environment}) and data (Section~\ref{sec:data}), and then differentiate between two pipelines: one for training and static data preparation (Section~\ref{sec:training-pipeline}), and the actual inference pipeline (Section~\ref{sec:inference-pipeline}).

\subsection{Environment}
    \label{sec:environment}

We study the task allocation framework in the vehicular combat game \emph{World of Tanks: HEAT}~\cite{wargaming_wot_heat_2026}, focusing specifically on the \emph{Conquest} game mode. In this mode, two teams of $10$ vehicles each compete for control of certain objectives on square maps that vary in size and urbanization. In the evaluated matches, only a subset of vehicles, ranging from 1-10, is controllable by the recommender, typically to fill empty or abandoned player slots. Depending on the map, between 3-5 capture objectives, referred to as \emph{bases}, are active and can be captured by entering their vicinity within a specific radius while no enemy vehicle is present. Vehicles differ in type, equipment, mobility, and combat characteristics. In addition to a primary attack, vehicles may use skills and abilities that affect their tactical value in a given situation. When destroyed, a vehicle respawns after $12$ seconds at one of several predefined spawn zones. For this paper, we always pick the geometrically closest spawn zone for the recommended target base.

We consider high-level task allocation only, assuming that low-level navigation and combat are handled by separate systems, as is the case in \emph{World of Tanks: HEAT}. Every $3$ seconds, the system receives a snapshot of the current battle state and generates one recommendation for each controllable bot. In \emph{Conquest}, this recommendation is an assignment to a base. Multiple bots may be assigned to the same base when this is strategically beneficial.

More specifically, let $\mathcal{T}=\{0,3,6,\dots\}$ denote the decision times in seconds. At each decision time $t\in\mathcal{T}$, the system observes a battle snapshot $x_t = \big(\mathcal{V}_t, \mathcal{B}_t, g_t\big)$ where $\mathcal{V}_t$ denotes the set of vehicle states for both teams, $\mathcal{B}_t$ is the set of active bases together with their dynamic states, and $g_t$ the remaining global game state such as score, map identity, and elapsed time. Let $\mathcal{C}_t \subseteq \mathcal{V}_t$ further denote the subset of controllable allied bots at time $t$, with cardinality $|\mathcal{C}_t| = n_t$, and let $\mathcal{Q}_t$ denote the set of currently actionable objectives. In the \emph{Conquest} mode, $\mathcal{Q}_t$ consists of the active bases. A recommendation step computes a mapping $\pi_t : \mathcal{C}_t \rightarrow \mathcal{Q}_t$ which assigns one high-level task to each bot; in our case, this task is a target base. As mentioned above, the mapping is many-to-one, i.e. several bots may target the same base.

The quality of an assignment depends on both local and global factors, including travel time, base state, unit force concentration, vehicle capabilities, and the assignment choices made for other controllable bots. The goal is, therefore, not to optimize each vehicle independently, but to produce a coordinated team-level assignment that improves downstream match performance, measured primarily by win rate, under real-time constraints. We thus interpret a recommendation as strategically better when it improves long-term battle outcome rather than only maximizing a short-term local quantity such as immediate proximity to a base.

\paragraph{Decision horizon and solver scope.}
We formulate each recommendation as a one-step allocation and recompute it from a new snapshot every $3$ seconds. This receding-horizon design incorporates changes in ownership, pressure, health, human movement, and previous assignments at the next decision step rather than predicting them through a multi-step search. We did not use model-predictive planning, Monte Carlo tree search, or similar methods because they require reliable transition, opponent, and human-behavior models in a setting with only partial controllability. The logged data contains realized trajectories but not dependable counterfactual outcomes for alternative joint assignments, while low-level movement and combat add further branching and uncertainty. In this setting, deeper planning could amplify model error and reduce inspectability. The minimum-cost flow layer instead solves the current allocation problem exactly under its linear scores and capacity constraints, while repeated re-optimization provides adaptation over time. That said, we do not claim that multi-step planning is unsuitable in general; evaluating it with a validated simulator is left for future work.

\begin{figure*}[t]
  \centering
  \includegraphics[width=0.68\linewidth]{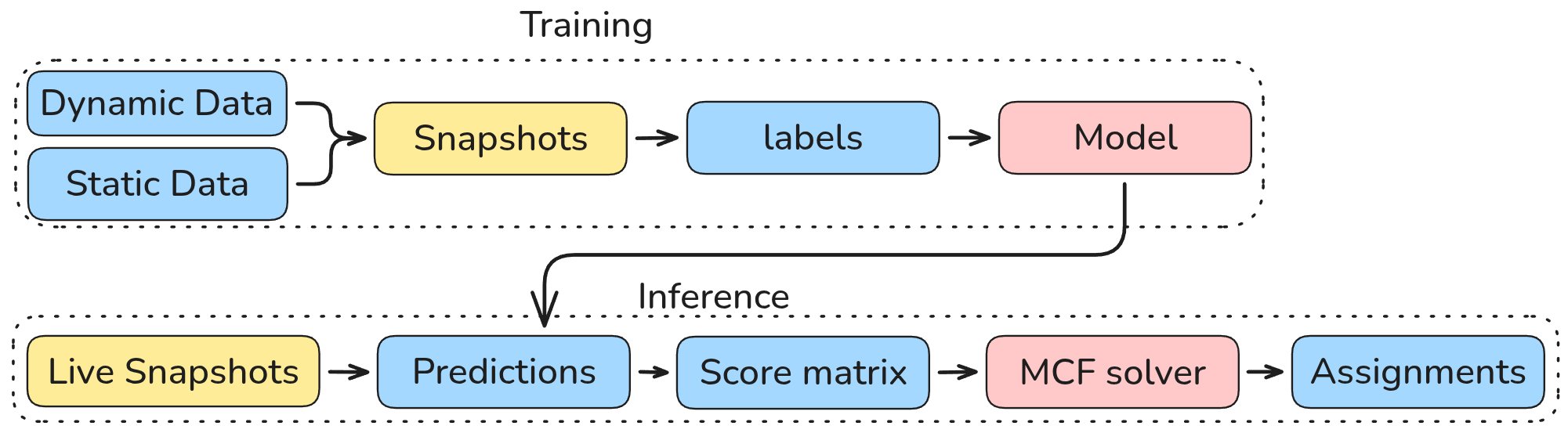}
  \caption{High-level training and inference pipeline. Offline gameplay logs and static data form labeled battle snapshots for training the prediction model. At runtime, live snapshots are used to predict tactical signals, which are merged with heuristics into a vehicle--objective score matrix, and optimized by a minimum-cost flow solver to recommend coordinated assignments.}
  \label{fig:pipeline}
\end{figure*}

\subsection{Data}
    \label{sec:data}

We first describe how gameplay data is converted into battle-state snapshots and then the training dataset.

\paragraph{Snapshot Construction:} Battle snapshots are constructed from two data sources:

\begin{description}[leftmargin=*]
    \item[Dynamic data] -- Battle events, match outcomes, player actions, game metrics, and other time-dependent values used to reconstruct battle states over time.
    \item[Static game data] -- Versioned map definitions, mechanics parameters, and vehicle configurations, including objective locations and radii, map-specific parameters, and vehicle characteristics.
\end{description}

\noindent Dynamic records are aligned with static data by game version and transformed into snapshot sequences. The resulting representation is used for both offline training and online inference, reducing training--inference skew by keeping gameplay events tied to correct map layout, objective configuration, and vehicle parameters. For supervised training, labels are assigned as described in Section~\ref{sec:label-generation}. At runtime, the inference service constructs state snapshots from live gameplay events using the same feature schema as described in Section~\ref{sec:inference-pipeline}. Static game data are precomputed for each supported game version, since maps, objectives, and vehicle parameters may change between versions.

\paragraph{Training Dataset:} The dataset contains approximately $22{,}500$ complete battles logged from the live game, each with $20$ vehicles and snapshots sampled every $3$ seconds, yielding on average $180$ snapshots per match. Training labels are constructed only from battles without bots, such that the tactical targets are derived from human gameplay rather than pre-existing bot behavior. At inference time, the model observes mixed human--bot matches but controls only the bot subset.

Only battles with recorded player events entered dataset construction. Snapshots required at least one valid active vehicle state and one applicable \emph{Conquest} label. These criteria favor active and observable states and may underrepresent spawn periods, inactive or disconnected vehicles, early exits, and tactically ambiguous situations. 

Each retained snapshot contains (i) dynamic per-vehicle state such as position, normalized health, death state, active effects, and distances to candidate bases; (ii) static vehicle metadata such as team, vehicle type, and normalized vehicle characteristics; and (iii) global and objective-level state such as elapsed time, score, map name, base owner and status, and capture progress. We include both the current and previous vehicle state to provide short-term temporal context while retaining snapshot-based inference. Recommendations are generated only for bots, while all vehicles still contribute to the observed tactical context.

To reduce leakage across nearby matches and across patch progression, we use a temporal split of $70\%/10\%/20\%$ for training, validation, and test data. All significance testing and outcome comparisons in the evaluation are reported at battle level, while snapshot-level metrics are used only for direct predictive evaluation where the target is defined at the snapshot granularity.

\subsection{Training Pipeline}
    \label{sec:training-pipeline}

This pipeline (see Figure~\ref{fig:pipeline}, top) spans data preparation through model fitting. Training requires converting each battle snapshot into weakly supervised targets for the tactical outcomes later predicted by the model.

\subsubsection{Label generation}
    \label{sec:label-generation}

As strategic intent is not directly observable from the available telemetry and is not explicitly labeled, we infer it from future trajectories and events. A monotonic decrease in distance to a base is used as a proxy for ``commitment'', starting when the vehicle first begins moving toward that base and ending when the base stops being contested. Because events can overlap, labels are weak and predictive heads are optimized under masked supervision. We construct multiple per-vehicle targets: 

\begin{itemize}
    \item \emph{will go to base / token $X$} (binary tokens head; a token is the base itself, so this head learns where human players typically go in the given situation),
    \item \emph{will be on the winning team} (masked binary head),
    \item \emph{own team will achieve local superiority at base $X$} (masked binary head),
    \item \emph{will survive the engagement at base $X$} (masked binary head),
    \item \emph{will deal damage at base $X$} (masked binary head),
    \item \emph{will receive damage at base $X$} (masked binary head),
    \item \emph{time in seconds to reach base $X$} (masked regression head),
    \item \emph{cumulative contribution to base $X$ by battle end} (masked regression head).
\end{itemize}

\begin{figure*}[t]
  \centering
  \includegraphics[width=0.65\linewidth]{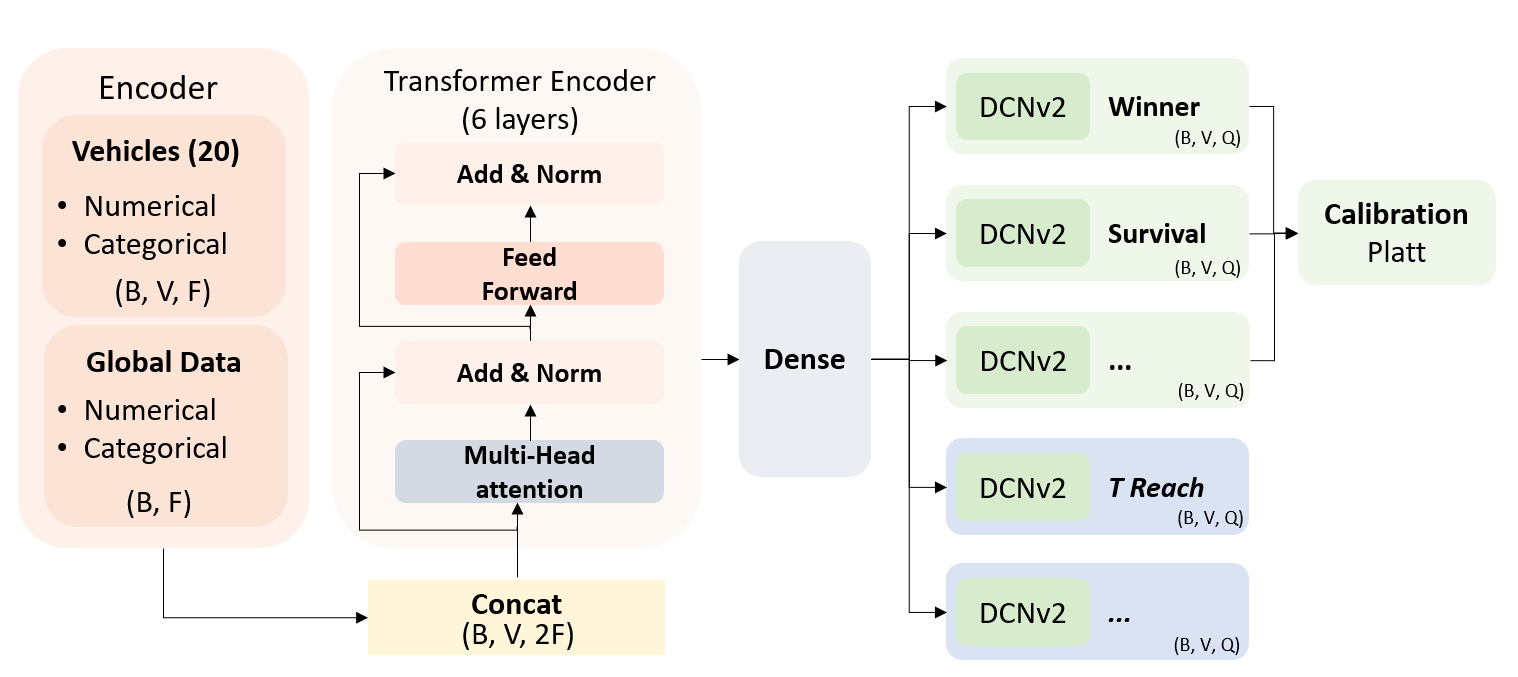}
  \caption{Model architecture used in the training pipeline. Vehicle and global data are embedded and concatenated into tokens, fed to a multi-head attention layer, and passed to a set of prediction heads.}
  \label{fig:model-architecture}
\end{figure*}

\subsubsection{Model}
A masked multi-head model, visualized in Figure~\ref{fig:model-architecture}, is trained to produce per-vehicle--query predictions used by the downstream optimizer described in Section~\ref{sec:inference-pipeline}. Let $B$ denote the batch size, $V$ the number of vehicles, and $Q$ the number of query objectives. Each prediction head outputs one scalar for each vehicle--query pair $(v,q)$, resulting in tensors of shape $B \times V \times Q \times 1$.

Numerical vehicle and global features are normalized with \emph{LayerNorm}~\cite{ba2016layernormalization} and concatenated with learned categorical embeddings. The global context embedding is replicated across vehicles, concatenated to vehicle embeddings, projected, and passed to a masked multi-head self-attention encoder~\cite{vaswani2023attentionneed}. Each attended vehicle token is linearly projected to the final embedding dimension and combined with each query embedding. For each vehicle--query pair, we form
\begin{equation}
  \mathbf{z}_{b,v,q}
  =
  \big[
  \mathbf{h}_{b,v}
  \,;\,
  \mathbf{W}_q \mathbf{q}_{b,q}
  \,;\,
  \mathbf{h}_{b,v} \odot \mathbf{W}_q \mathbf{q}_{b,q}
  \big],
\end{equation}
where $\mathbf{h}_{b,v}$ is the vehicle embedding, $\mathbf{q}_{b,q}$ is the query embedding, $\mathbf{W}_q$ projects queries to the vehicle-embedding space, and $\odot$ denotes element-wise multiplication. The resulting pairwise feature combines the vehicle state, the queried objective, and their multiplicative interaction. Each pairwise feature is processed by a task-specific \emph{Deep \& Cross Network v2} (DCNv2) head~\cite{Wang_2021}. The head contains parallel cross and deep branches. Its cross-branch applies $L$ residual
matrix cross-layers,
\begin{equation}
\mathbf{x}_{l+1}
=
\mathbf{x}_0 \odot
\left(\mathbf{W}_l\mathbf{x}_l+\mathbf{b}_l\right)
+\mathbf{x}_l,
\qquad
\mathbf{x}_0=\mathbf{z}_{b,v,q},
\label{eq:dcnv2-cross-layer}
\end{equation}
while the deep branch applies a multilayer perceptron to $\mathbf{z}_{b,v,q}$. The outputs of both branches are concatenated and linearly projected to produce either a binary logit or a regression value for the corresponding prediction head.
We train with \emph{PyTorch AdamW}~\cite{loshchilov2019decoupledweightdecayregularization}, cosine annealing~\cite{loshchilov2017sgdrstochasticgradientdescent}, batch size 256, gradient clipping, bfloat16 precision, \emph{FAMO} multi-task balancing~\cite{liu2023famofastadaptivemultitask}, and early stopping.

\paragraph{Model capacity.}
The architecture was chosen as a practical deployment trade-off, not as an optimal model size. Given the small number of vehicle and objective tokens, we prioritized a compact model that could be retrained efficiently and run on a CPU well within the $3$-second decision interval. We did not perform a scaling study and therefore make no claim that larger models would not improve offline metrics; however, such gains alone would not demonstrate better assignments or higher win rate.

The supervision is partial because logged replays contain only one realized trajectory. We only observe the outcome for the base that the player actually attempted to contest or support, but not the counterfactual outcomes for alternative assignments. Losses are therefore computed only for valid vehicle--query labels, motivating local tactical estimators rather than an end-to-end assignment policy.

\paragraph{Masked Training Objectives.}
Because replay data reveal outcomes only for realized vehicle--objective pairs, each prediction head is trained only on valid labels. For a binary head, let $z_{bvq}$ denote the predicted logit, $y_{bvq}\in\{0,1\}$ the target, and $m_{bvq}\in\{0,1\}$ the validity mask. We minimize class-weighted binary cross-entropy,
\begin{equation}
    \begin{aligned}
    \mathcal{L}_{\mathrm{bin}} ={}& -\frac{1}{N_{\mathrm{bin}}}
    \sum_{b,v,q} m_{bvq}\big[w^{+}y_{bvq}\log \sigma(z_{bvq}) \\
    &+(1-y_{bvq})\log(1-\sigma(z_{bvq}))\big],
    \end{aligned}
    \label{eq:binary-loss}
\end{equation}
where $N_{\mathrm{bin}}=\sum_{b,v,q}m_{bvq}$ and $w^{+}$ compensates for class imbalance. Binary outputs are calibrated using Platt scaling~\cite{platt2000probabilistic} fitted on the validation split before being used as probabilities by the downstream optimizer.

For a regression head with prediction $\hat{r}_{bvq}$ and target $r_{bvq}$, we minimize the masked \emph{SmoothL1} loss~\cite{girshick2015fastrcnn},
\begin{equation}
    \mathcal{L}_{\mathrm{reg}}
    =
    \frac{1}{N_{\mathrm{reg}}}
    \sum_{b,v,q}
    m_{bvq}\,
    \operatorname{SmoothL1}\!\left(\hat{r}_{bvq},r_{bvq}\right),
    \label{eq:regression-loss}
\end{equation}
where $N_{\mathrm{reg}}$ is the number of valid regression labels.

\subsection{Inference Pipeline}
    \label{sec:inference-pipeline}

\sloppy{The inference pipeline (Figure~\ref{fig:pipeline}, bottom) converts each battle snapshot into a constrained assignment problem. Since only a subset of allied vehicles is controllable and only a subset of bases is active at a given time, all computations are performed on the sliced score matrix $\texttt{scores}[\texttt{controllable\_vehicles}, \texttt{active\_bases}].$} The matrix is assembled from model predictions and hand-designed tactical terms, then passed to the assignment solver. The main components are:

\begin{itemize}
    \item \textbf{Model prediction.}
    The trained model produces per-vehicle--base outputs. Calibrated binary heads are used as probabilities, while regression heads are normalized to $[0,1]$ and combined with probability-valued terms.

    \item \textbf{Distance and travel cost.}
    Traveling to a base is strategically costly and is therefore penalized. The distance term is computed from the Euclidean distance between the vehicle and the base. Both raw geometric distance and predicted time-to-reach may contribute to the final score, allowing the system to penalize objectives that are either far away or slow to reach because of terrain and contextual effects.

    \item \textbf{Force balance.}
    Local force advantage at a base is approximated by a heuristic we termed \emph{superiority}. This heuristic aggregates nearby friendly and enemy units into a bounded force-balance estimate. It depends on combat-relevant vehicle attributes and health within a buffer zone around the base. It expresses whether assigning an additional bot to that region is tactically supported by the local team composition.

    \item \textbf{Assignment stickiness.}
    To prevent frequent reassignment, a stickiness penalty is applied to the score matrix. If a vehicle was assigned to base $q$ in the previous recommendation step, all alternative bases $q' \neq q$ are penalized.
\end{itemize}

\noindent Assignments are computed using a min-cost flow (\emph{MCF}) solver~\cite{google_ortools_mincostflow} under game-specific constraints. These include base-dependent slot counts, which allow more bots to be assigned to important or pressured bases, and soft minimum-garrison incentives, i.e., bonuses that make assigning at least one bot to selected defensive bases cheaper rather than enforcing a hard lower bound. Human occupancy reduces the effective bot capacity before the flow problem is solved, thereby preventing unnecessary overcommitment to already covered objectives.

Before optimization, a policy layer computes adaptive \emph{attack slots} for each active base from base-level tactical signals, including enemy pressure, strategic priority, contested ownership, and current human coverage. For each controllable vehicle $v$ and active base $q$, the final assignment score is computed as
\begin{equation}
\begin{aligned}
    S_{vq} =\;&
    w_{\text{tokens}}\, p_{\text{tokens}} +
    w_{\text{winner}}\, p_{\text{winner}} \\
    &+w_{\text{superiority}}\, p_{\text{superiority}} +
    w_{\text{survival}}\, p_{\text{survival}} \\
    &+w_{\text{contribution}}\, p_{\text{contribution}} +
    w_{\text{damage}}\, p_{\text{damage}} \\
    &-w_{\text{hurt}}\, p_{\text{hurt}} +
    w_{\text{balance}}\, \tilde{b} \\
    &-w_{\text{distance}}\, \tilde{d} -
    w_{\text{reach}}\, \tilde{t}_{\text{reach}} -
    P_{\text{stickiness}},
\end{aligned}
\end{equation}
where the individual terms combine calibrated model predictions, normalized regression outputs, heuristics and rule-based bonuses. Their weights are manually chosen rather than learned end-to-end, and were iteratively adjusted based on player feedback, game-designer expectations, and observed metrics. As a result, they reflect design judgments and may bias assignments toward the behaviors the team focused on. We do not claim these weights are optimal, transferable, or independently validated; their main value is making the trade-offs explicit and easy to adjust. 

\begin{figure*}[t]
  \centering
  \includegraphics[width=0.9\linewidth]{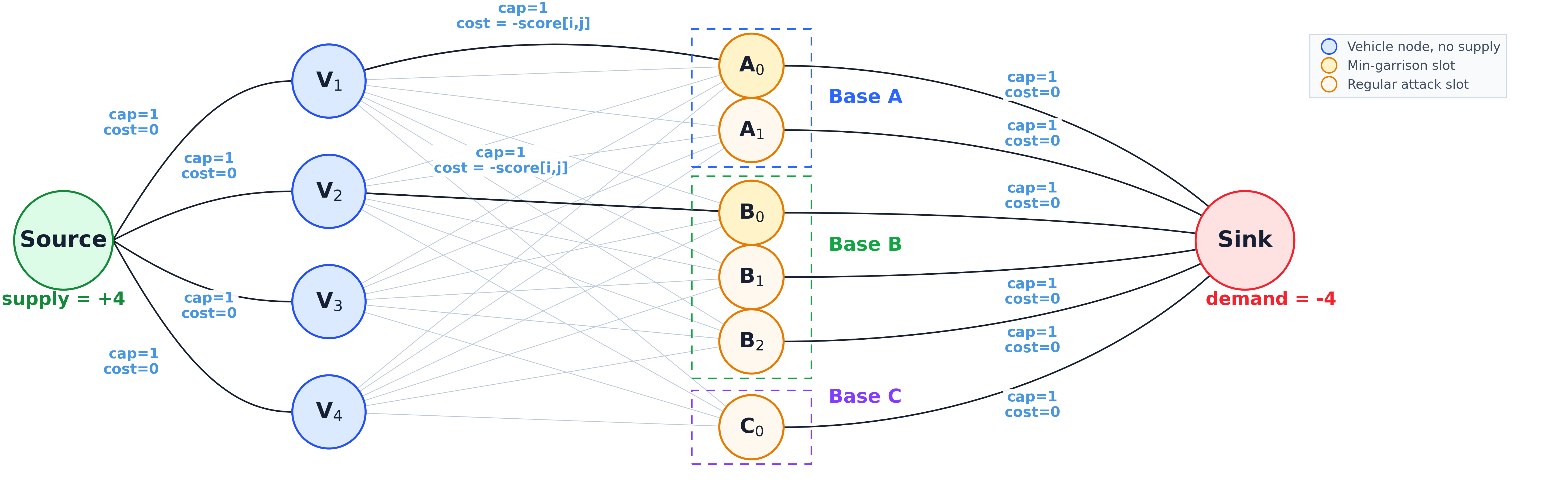}
  \caption{MCF formulation for coordinated task assignment. Each controllable vehicle sends one flow unit to an objective slot encoding base capacities and optional minimum-allocation incentives; edge costs use negated assignment scores.}
  \label{fig:min-cost-flow}
\end{figure*}

The MCF graph, visualized in Figure~\ref{fig:min-cost-flow}, contains a source node, one node per controllable vehicle, one or more slot nodes per active base, as well as a sink node. The source supplies $n_t$ flow units, and unit-capacity source--vehicle edges ensure that each vehicle may send at most one flow unit to a base slot. Base slots encode capacity constraints and optional minimum-garrison incentives. Edge costs are obtained from the negated assignment scores after scaling to integer costs for the solver. If the set of active bases is empty, or if the solver cannot assign all bots because of slot limits, the remaining bots retain their previous assignment or fall back to a hold/default behavior. In these cases, a complete mapping $\pi_t$ is not guaranteed.

\begin{figure}[ht]
  \centering
  \includegraphics[
  width=1.0\linewidth,
  trim=0 100pt 0 100pt,
  clip
]{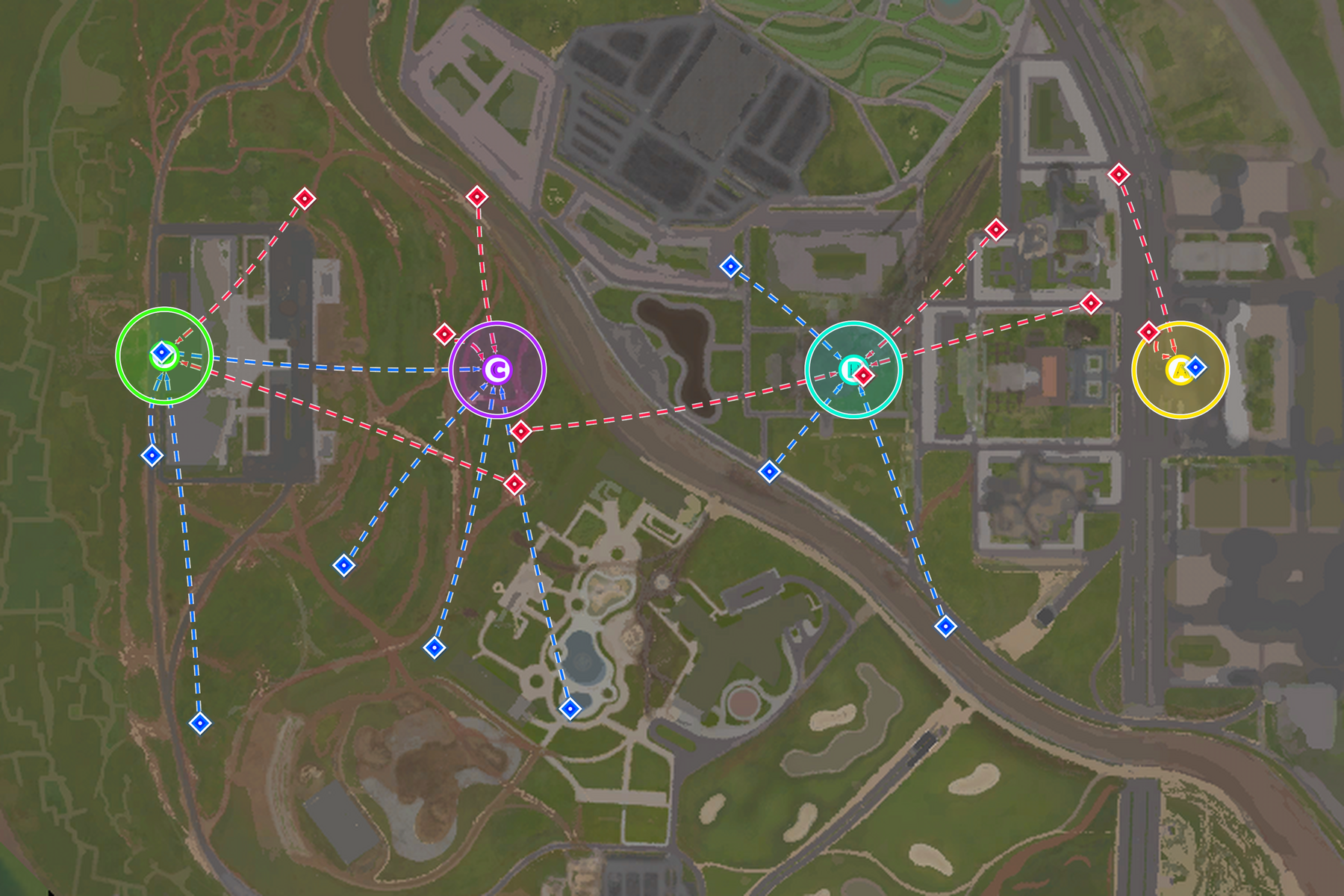}
  \caption{Spatial assignment view for both bot teams at one match state. Colored circles mark active base regions, and stippled blue/red lines connect each team's bots to their assigned objectives, showing how the solver distributes bots across bases in the current tactical state.}
  \label{fig:example-assignment}
\end{figure}

\begin{figure*}[ht!]
  \centering
  \includegraphics[
    width=0.8\linewidth,
    height=5.25cm,
    trim={0 0.5cm 0 0},
    clip
  ]{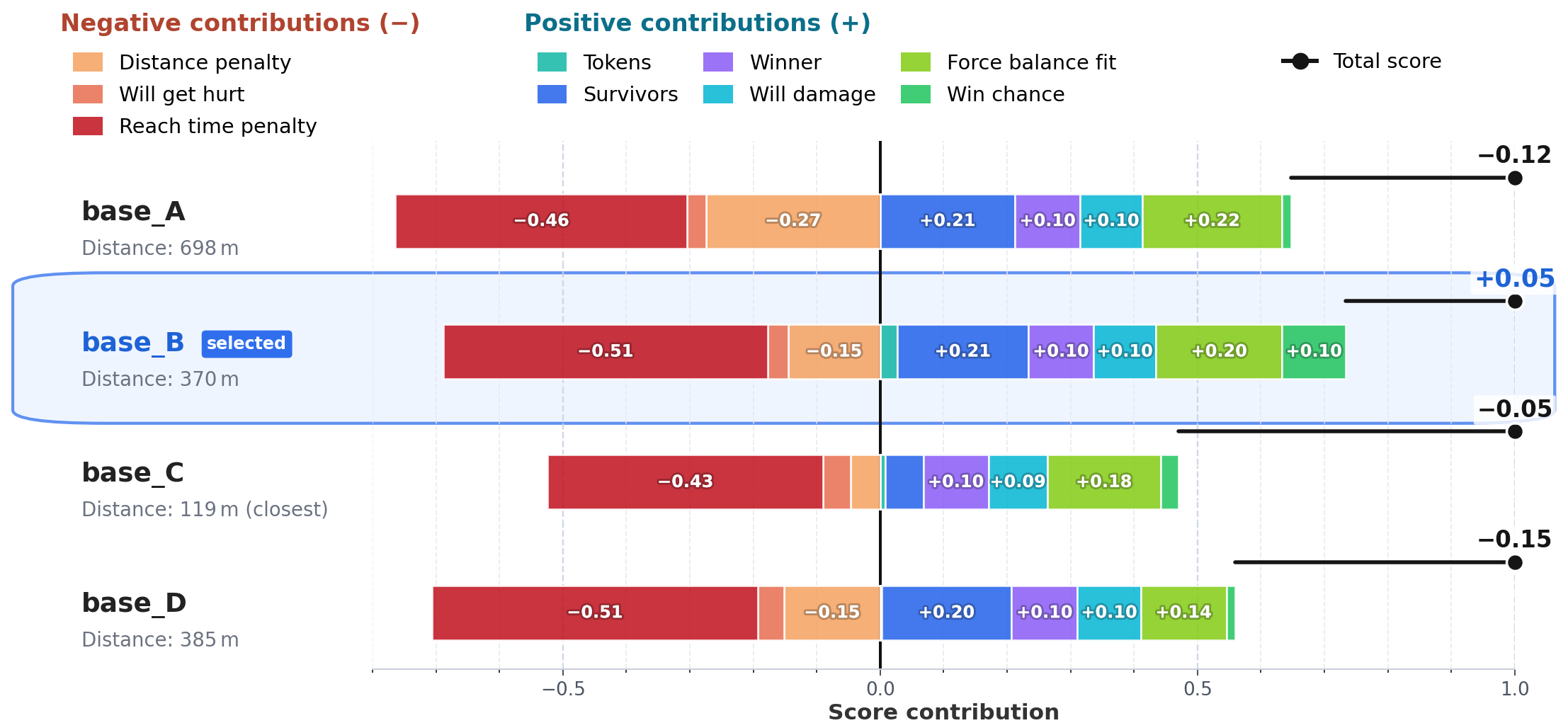}
   \caption{Assignment-score decomposition for a bot at one decision step. For each candidate base, colored bars show individual score-component contributions, and the black marker shows the final aggregated score. Here, Base~C is closest, but Base~B scores highest because tactical context outweighs the distance advantage.}
  \label{fig:example-assignment-scores}
\end{figure*}

Lastly, to allow game AI designers to better understand the outputs of the model, a visualization tool -- built on the above inference pipeline -- lets them inspect selected recommendations in their spatial context, as shown in Figure~\ref{fig:example-assignment}. It also shows score decompositions that explain whether a recommendation was mainly driven by learned model output, heuristic terms, or solver constraints, as shown in Figure~\ref{fig:example-assignment-scores}. These views are meant for manual inspection and tuning, rather than for being used for a quantitative evaluation of interpretability.

\section{\uppercase{Evaluation}}
    \label{sec:evaluation}

We evaluate the proposed hybrid task allocation framework with respect to four system requirements: useful local tactical estimates, coordinated and stable team-level assignments, improved battle-level outcomes, and low-latency inference. We address these requirements through offline predictive evaluation, offline assignment diagnostics, online A/B evaluation, and latency measurements.

\subsection{Offline Predictive Evaluation}

We first evaluate the multi-head model independently of the assignment solver. This measures whether the learned heads provide useful tactical estimates before they are consumed by the downstream optimizer. For binary heads, AUC (area under the curve), precision, and recall are reported on the held-out test split, while expected calibration error (ECE) is reported on the validation split used to fit Platt scaling (Table~\ref{tab:model-metrics}). For regression heads, MAE (mean absolute error) and RMSE (root mean square error) are reported on the held-out test split (Table~\ref{tab:model-metrics-reg}).

Based on the heads introduced in Section~\ref{sec:label-generation} and reported in Table~\ref{tab:model-metrics-all}, the \emph{survival}, \emph{damage}, \emph{hurt}, and \emph{superiority} heads indicate that the model can distinguish relevant tactical outcomes substantially better than chance. \emph{Tokens} is a strong discriminator, but with low precision due to the high class imbalance of this specific head. The \emph{winner} head is weaker, which is expected because the final outcome depends on longer-horizon interactions. The low ECE values suggest that the predicted probabilities can be used as meaningful soft costs rather than only as ranking scores. The regression heads show low errors for both \emph{time-to-reach} and \emph{final contribution}. Overall, the heads are sufficiently informative as local estimates to serve as inputs to the optimizer.

\subsection{Offline Assignment Evaluation}

To evaluate offline assignment, we compare the following variants of the method:
\begin{description}
    \item[\textsc{Greedy-model}:] uses learned model scores with slot capacities; bots are greedily assigned to their highest-scoring base with remaining capacity.
    \item[\textsc{Solver-model-only}:] uses the MCF solver with learned model scores only.
    \item[\textsc{Full}:] uses the complete score formulation and constrained solver.
\end{description}

\begin{table}[t]
    \caption{Predictive performance of model heads on the held-out split. Bootstrap 95\% confidence intervals are shown as subscript/superscript bounds. ECE values are post-calibration point estimates on the held-out validation split used to fit Platt scaling. Final contribution is unitless and normalized to [0,1] within each battle.}
    \label{tab:model-metrics-all}
    \small
    \begin{subtable}[t]{1.0\columnwidth}
    \setlength{\tabcolsep}{4pt}
    \centering
    \caption{Binary heads}
    \begin{tabular}{lcccc}
        \toprule
        Head & AUC & Precision & Recall & ECE \\
        \midrule
            Winner        & $0.697_{0.684}^{0.710}$ & $0.666_{0.655}^{0.678}$ & $0.662_{0.650}^{0.674}$ & 0.0156 \\[2pt]
            Superiority   & $0.786_{0.780}^{0.790}$ & $0.721_{0.717}^{0.726}$ & $0.717_{0.713}^{0.721}$ & 0.0115 \\[2pt]
            Survival      & $0.801_{0.799}^{0.802}$ & $0.790_{0.788}^{0.792}$ & $0.712_{0.708}^{0.716}$ & 0.0042 \\[2pt]
            Damage        & $0.798_{0.796}^{0.800}$ & $0.916_{0.914}^{0.917}$ & $0.792_{0.787}^{0.796}$ & 0.0146 \\[2pt]
            Hurt          & $0.797_{0.795}^{0.798}$ & $0.897_{0.895}^{0.898}$ & $0.726_{0.722}^{0.730}$ & 0.0135 \\[2pt]
            Tokens        & $0.957_{0.956}^{0.957}$ & $0.459_{0.456}^{0.461}$ & $0.946_{0.945}^{0.946}$ & 0.0068 \\
        \bottomrule
    \end{tabular}
    \label{tab:model-metrics}
    \end{subtable}\\
    
    \begin{subtable}[t]{1.0\columnwidth}
    \setlength{\tabcolsep}{12pt}
    \centering
    \caption{Regression heads}
    \small
    \setlength{\tabcolsep}{3pt}
    \begin{tabular}{lcc}
        \toprule
        Head & MAE & RMSE \\
        \midrule
            Time-to-reach (s)  & $6.427_{6.394}^{6.464}$ & $9.871_{9.834}^{9.908}$ \\[2pt]
            Final contribution & $0.173_{0.171}^{0.174}$ & $0.210_{0.209}^{0.212}$ \\
        \bottomrule
    \end{tabular}
    \label{tab:model-metrics-reg}
    \end{subtable}
\end{table}

We evaluate assignment quality on held-out replay snapshots using diagnostic metrics computed over the observed replay continuation. Since replay data contains only the trajectory that actually occurred, these metrics do not estimate counterfactual policy value. Instead, they characterize historical-state behavior: selected objectives, whether observed vehicles reached them, recommendation stability, and replay outcomes associated with the selected objectives. We report five diagnostics using a 60-second replay horizon where applicable (first aggregated per battle and then averaged across battles): 
\begin{itemize}
    \item \emph{Capture conversion}: whether assignments to neutral or enemy-controlled bases are followed by allied control within the horizon.
    \item \emph{Defensive retention}: whether assigned allied-controlled bases remain allied-controlled throughout the horizon.
    \item \emph{Coverage}: the fraction of critical bases receiving at least one controllable assignment, where critical means enemy-controlled, neutral, contested, or allied-controlled under nearby enemy pressure.
    \item \emph{Switches per bot per minute}: target changes between consecutive valid recommendations.
    \item \emph{Reach@60}: whether the vehicle's observed replay trajectory, which was not generated under the recommender, enters the assigned base radius within 60 seconds.
\end{itemize}

\begin{table*}[t]
\centering
\caption{Offline assignment diagnostics on held-out replay snapshots. Lower is better for switches. Values report means with 95\% battle-level bootstrap confidence intervals over the test battles.}
\label{tab:assignment-metrics}
\small
\begin{tabularx}{1.0\textwidth}{X>{\centering}p{2cm}>{\centering}p{2cm}>{\centering}p{2cm}>{\centering}p{2cm}>{\centering\arraybackslash}p{2cm}}
\toprule
Method 
& Capture conversion 
& Defensive retention 
& Coverage 
& Switches [bot/min] 
& Reach@60 \\
\midrule
\textsc{Greedy-model}
& $\mathbf{0.315}_{0.314}^{0.317}$
& $0.350_{0.347}^{0.353}$
& $\mathbf{0.965}_{0.965}^{0.966}$
& $1.692_{1.688}^{1.697}$
& $0.156_{0.155}^{0.157}$ \\[2pt]
\textsc{Solver-model-only}
& $0.315_{0.314}^{0.316}$
& $\mathbf{0.351}_{0.348}^{0.354}$
& $0.964_{0.963}^{0.965}$
& $1.742_{1.738}^{1.746}$
& $0.181_{0.180}^{0.182}$ \\[2pt]
\textsc{Full}
& $0.254_{0.253}^{0.255}$
& $0.304_{0.301}^{0.307}$
& $0.867_{0.864}^{0.870}$
& $\mathbf{1.443}_{1.436}^{1.450}$
& $\mathbf{0.371}_{0.368}^{0.373}$ \\
\bottomrule
\end{tabularx}
\end{table*}

\noindent As seen from Table~\ref{tab:assignment-metrics}, \textsc{Full} has the strongest offline effects on trajectory alignment and stability. It improves \emph{Reach@60} over both ablations, indicating that its score formulation selects objectives more consistent with observed near-future trajectories. It also reduces assignment switching relative to both ablations. The lower \emph{capture conversion} and \emph{defensive retention} should be interpreted as difficulty-sensitive replay outcomes, not direct performance regressions. These metrics are conditional on the objectives selected by the policy. The \textsc{Greedy-model} and \textsc{Solver-model-only} variants assign more broadly and often favor locally attractive objectives that are easier to capture, defend, or retain. In contrast, the \textsc{Full} policy uses heuristic terms and dynamic assignment constraints to prioritize pressured, contested, or tactically urgent objectives. The \emph{coverage} result reflects the same breadth--depth trade-off. The \textsc{Greedy-model} and the \textsc{Solver-model-only} variants cover nearly all critical bases, while \textsc{Full} focuses controllable vehicles on fewer, higher-priority objectives. 

\subsection{Battle-Level Online Evaluation}

To evaluate whether assignment changes translate into battle outcomes, we conducted three separate online experiments against the closest-base baseline, one per evaluated policy. The analyzed datasets contained $400$ battles for \textsc{Greedy-model}, $400$ for \textsc{Solver-model-only}, and $410$ for \textsc{Full}. Each policy was compared with the closest-base control group within the same battles, with bot-level outcomes aggregated within each group to yield one paired observation per battle. Policy and baseline bots were on opposing teams, so lift equals $2p-1$ for win rate $p$.

\begin{table}[t]
\centering
\caption{Battle-level online A/B evaluation against the closest-base baseline. Policy wins were $263/400$, $254/400$, and $281/410$, respectively; no draws occurred. Lift is the raw win-rate difference in percentage points, calculated before rounding. Wilson 95\% confidence intervals are shown as sub-/superscripts. Two-sided $p$-values use paired $t$-tests of within-battle outcomes without CUPED adjustment.}
\label{tab:ab-results}
\small
\setlength{\tabcolsep}{2pt}
\begin{tabular}{@{}lccc@{}}
\toprule
Policy & Win rate & Lift & $p$-value\\
\midrule
\textsc{Greedy-model}      & $65.8_{61.0}^{70.2}$\% & $+31.5_{21.9}^{40.5}$\,pp & $<.001$\\
\textsc{Solver-model-only} & $63.5_{58.7}^{68.1}$\% & $+27.0_{17.3}^{36.1}$\,pp & $<.001$\\
\textsc{Full}              & $68.5_{63.9}^{72.8}$\% & $+37.1_{27.8}^{45.7}$\,pp & $<.001$\\
\bottomrule
\end{tabular}
\end{table}

As shown in Table~\ref{tab:ab-results}, all three evaluated variants achieved significantly higher win rates than their corresponding closest-base controls. The \textsc{Full} method achieved the highest observed win rate, $68.5\%$, and the largest lift, $37.1$ percentage points.
The observed win rate of the \textsc{Greedy-model} was slightly higher than that of the \textsc{Solver-model-only}. However, each variant was evaluated in a separate experiment against its own control, and these tests do not establish statistically significant differences between the evaluated variants or isolate the contribution of MCF. The improvements over the closest-base controls provide evidence that the proposed recommendations benefit final battle outcomes, extending the evaluation beyond local proximity and immediate objective attraction.

\subsection{Latency Evaluation}

Finally, we evaluate runtime latency. As recommendations are generated every 3 seconds (see Section~\ref{sec:environment}), the \textsc{Full} pipeline must remain well below this interval. We measure single-snapshot latency using ONNX Runtime CPU inference on an Intel Core i9-13900HX processor. Timings exclude external transport, network delay, and game-server communication. Server timings may differ, but the ${\sim}100\times$ margin leaves ample headroom. As evident from Table~\ref{tab:latency}, the \textsc{Full} pipeline remains far below the 3-second interval, with a p99 latency of $28.12$ ms. Runtime is dominated by model inference, while score assembly and assignment solving remain small because the optimization problem contains at most 10 controllable vehicles and 3 to 5 active bases. These results show that the structured optimization layer does not introduce a production-relevant bottleneck.

\begin{table}[t]
    \centering
    \caption{Single-snapshot latency (ms), ONNX Runtime on CPU.}
    \label{tab:latency}
    \small
    \begin{tabular}{lccc}
    \toprule
    Stage & Median & p95 & p99 \\
    \midrule
    Feature construction    & 0.31  & 0.37  & 0.39 \\
    Model inference         & 16.57 & 17.16 & 17.54 \\
    Score + solver          & 5.69  & 6.73  & 11.63 \\
    Total                   & 22.56 & 24.05 & 28.12 \\
    \bottomrule
    \end{tabular}
\end{table}

\section{\uppercase{Discussion}}

Our results support the choice of separating tactical prediction from global assignment optimization. The learned heads provide useful short-horizon signals, while the constrained solver converts these local estimates into coordinated team-level assignments. The weaker \emph{winner} head indicates that final objective success is harder to infer from short temporal context than immediate tactical outcomes. This advocates using the model as a tactical scorer rather than as a direct strategic outcome predictor, consistent with predict-then-optimize work that judges predictions by the decisions they induce~\cite{elmachtoub2022spo}.

Offline assignment results show that coordination and robustness matter beyond local score maximization. Compared with both ablations, the \textsc{Full} method produces fewer assignment switches and recommendations that are more consistent with observed vehicle movements, reducing wasted travel time and indecisiveness. However, the lower \emph{capture conversion} and \emph{defensive retention} scores of the \textsc{Full} method should be interpreted carefully. Policies that select easier or already favorable objectives may score higher offline without being strategically better. The \textsc{Full} method instead prioritizes pressured or tactically urgent objectives, showing limits of offline diagnostics for estimating counterfactual policy value.

The online results clarify the ablations. The \textsc{Greedy-model} is a slot-capacity greedy policy: bots are processed by assignment confidence and assigned to their highest-scoring base with capacity. Its higher observed win rate than the \textsc{Solver-model-only} may indicate, though without statistical support, that learned scores alone are insufficient as a global MCF objective. If scores are noisy or poorly calibrated, the solver can optimize an imperfect objective. The \textsc{Full} method achieves the highest win rate, suggesting that constrained optimization is most useful with the complete strategic score formulation.

The min-cost flow layer converts local utilities into globally feasible assignments under base capacities, human occupancy constraints, and many-to-one allocations~\cite{bertsekas1998network,gerkey2004formal,korsah2013taxonomy}. This makes the policy more coordinated than independent per-bot assignments and more controllable than a fully learned policy. The spatial-assignment and score-decomposition views further make recommendations inspectable and tunable.

We did not use reinforcement learning because the available data are observational and provide neither reliable counterfactual feedback nor a safe environment for online exploration. In addition, explicit constraints on capacity, occupancy, and assignment stability are easier to enforce and inspect with the chosen hybrid design. We therefore treat RL as a possible extension rather than a necessary baseline here.

Finally, the systems results show that the approach is compatible with real-time production constraints. Overall, the findings suggest that learning-assisted constrained optimization is a practical middle ground between hand-authored heuristics and end-to-end learned models for contested-objective game AI.

These findings apply to the evaluated setting. We do not claim global optimality or that the chosen model size and score weights are uniquely best. Offline diagnostics describe historical behavior, while the online A/B test supports comparison with the tested closest-base baseline. Generalization to other modes, maps, populations, or real-world systems remains future work.

\section{\uppercase{Limitations and Future Work}}

Several limitations remain. First, our training signals are weak and trajectory-derived, so the original intent is unknown and only inferred from observed behavior. This heuristic can be noisy in situations involving mid-trajectory re-planning, long-range actions, detours, or repositioning not captured as individual training labels. In addition, this approach cannot produce real counterfactual outcomes and depends on masked learning to predict those. The intent proxy was not validated, training excludes bots -- unlike inference -- and ECE uses the calibration split.

Another limitation is the use of static weights in the score matrix. While these weights make bot behavior transparent and manually tunable, a more dynamic and adaptive weighting scheme could further optimize performance. Human behavior is also modeled only through occupancy estimation: bots currently cannot detect or adapt to complex human strategies or tactics, nor can they communicate with or receive commands from allied players.

This paper focused on the \emph{Conquest} mode of \emph{World of Tanks: HEAT}, so generalization to other modes, action spaces, and games remains to be tested. Future evaluations should stratify by map and player skill, study generalization across maps and game versions, and include A/B-test metrics beyond win rate such as behavioral statistics, churn, fairness, and perceived quality of experience.

\section{\uppercase{Conclusions}}

This paper studied strategic task allocation for contested objectives in adversarial multi-agent environments with partial controllability. It applied a hybrid heuristic and model-based predict-then-optimize, rule-constrained pipeline for bot recommendations and evaluated it in a commercial game environment using both offline and online evaluation methods. Results show that learned prediction heads provide useful tactical signals, and the full assignment method produces more stable recommendations than simpler variants and higher win rates than the closest-base baseline. Overall, the work shows that learning-assisted constrained optimization can improve strategic coordination in production game AI while retaining deployment-critical robustness, interpretability, and controllability.

\bibliographystyle{apalike}
{\small
\bibliography{references}}

\end{document}